\documentclass[11pt]{article}
\pdfoutput=1
\usepackage{jcapmod}

\usepackage{booktabs}
\usepackage[english]{babel}
\usepackage{amsmath, amsfonts, amssymb, amsbsy, amstext, amsthm}
\usepackage{hyperref}
\usepackage{graphicx}
\usepackage{siunitx}
\usepackage{comment}
\usepackage[caption=false]{subfig}
\usepackage[utf8]{inputenc}
\allowdisplaybreaks[1]

\numberwithin{equation}{section}

\def\beq{\begin{equation}}
\def\eeq{\end{equation}}
\def\bea{\begin{eqnarray}}
\def\eea{\end{eqnarray}}

\def\Neff{N_{\rm eff}}

\def\md{m_{\rm dm}}

\DeclareRobustCommand{\SkipTocEntry}[4]{}

\usepackage{colortbl}
\definecolor{green2}{cmyk}{0, 1, 0.5, 0}

\begin{document}

\pagenumbering{roman}
\begin{titlepage}
\baselineskip=15.5pt \thispagestyle{empty}

\bigskip\

\vspace{1cm}
\begin{center}
{\fontsize{20.74}{24}\selectfont \sffamily \bfseries The Cosmology of Sub-MeV Dark Matter}
\end{center}

\vspace{0.2cm}
\begin{center}
{\fontsize{12}{30}\selectfont Daniel Green$^{\blacklozenge}$ and Surjeet Rajendran$^{\blacklozenge}$} 
\end{center}

\begin{center}
\vskip8pt

\textsl{$^\blacklozenge$ Department of Physics, University of California, Berkeley, CA 94720, USA}

\end{center}

\vspace{1.2cm}
\hrule \vspace{0.3cm}
\noindent {\sffamily \bfseries Abstract}\\[0.1cm]
Light dark matter is a compelling experimental target in light of stringent constraints on heavier WIMPs.  However, for a sub-MeV WIMP, the universe is sufficiently well understood at temperatures below 10 MeV that there is no room for it to be a thermal relic.  Avoiding thermalization is itself a strong constraint with significant implications for direct detection.  In this paper, we explore the space of models of sub-MeV dark matter with viable cosmologies.  We discuss several representative models chosen to have large cross-sections in direct detection experiments.  The parameter space of these models that is also consistent with astrophysical and lab-based limits is highly restricted for couplings to electrons but somewhat less constrained for nuclei.  We find that achieving nuclear cross-sections well-above the neutrino floor necessarily predicts a new contribution to the effective number of neutrino species, $\Delta \Neff = 0.09$ that will be tested by the next generation of CMB observations.  On the other hand, models with absorption signatures of dark matter are less restricted by cosmology even with future observations.

\vskip10pt
\hrule
\vskip10pt

\vspace{0.6cm}
\end{titlepage}

\thispagestyle{empty}
\setcounter{page}{2}
\tableofcontents

\clearpage
\pagenumbering{arabic}
\setcounter{page}{1}

\section{Introduction}
\label{sec:intro}

The identity of the dark matter is one of the great mysteries in physics.  Direct detection experiments to date have been particularly effective in the search for the classic models of WIMP dark matter with masses in the 1 GeV to 10 TeV range.   Null results from a number of experiments~\cite{Aprile:2012nq,Angloher:2015ewa,Agnese:2015nto,Tan:2016zwf,Akerib:2016vxi} rule out large regions of parameter space that were previously compatible with this simple picture.  The lack of evidence for dark matter in this mass range has motivated, in part, the search for dark matter at lower masses and/or with different detection signatures~\cite{Alexander:2016aln}.

Decreasing the mass of the dark matter can evade many conventional direct searches by lowering the recoil energy.  One might therefore imagine sub-MeV dark matter is weakly constrained down to the warm dark matter limit of ${\cal O}(10)$ keV~\cite{Viel:2013apy} and requires dedicated searches or new experiments.  However, the universe becomes transparent to neutrinos at temperatures below a few MeV and undergoes big bang nucleosynthesis (BBN) at temperatures of hundreds of keV.  Measurements of the neutrino energy density in the cosmic microwave background (CMB), via $\Neff$, and primordial abundances strongly constrain physics at times when a thermal sub-MeV mass particle is relativistic.  In fact, if the dark matter is a relic of thermal equilibrium with the Standard model, this low mass regime is already excluded by observations~\cite{Boehm:2012gr,Ho:2012ug,Steigman:2013yua,Boehm:2013jpa,Nollett:2013pwa,Steigman:2014uqa,Nollett:2014lwa}.  These exclusions are independent of the stringent constraints on s-wave dark matter annihilation~\cite{Boehm:2002yz, Boehm:2003bt,Padmanabhan:2005es, Ade:2015xua} and do not depend on the nature of the interactions that thermalize the dark matter.

Cosmological bounds are, of course, sensitive to assumptions about the physics of the early universe.  Our inference of physics at MeV temperatures is somewhat indirect and one might imagine that the limits on light dark matter are highly model-dependent.  However, dark matter is necessarily of cosmological origin and simply neglecting the cosmological constraints is not a viable alternative to the model-dependence.  The abundance of dark matter may arise from a non-thermal mechanism or even as an initial condition set by (or before) reheating.  However, for scenarios to be viable, we must ensure that it is sufficiently weakly coupled to the Standard Model to have never been in equilibrium.  Otherwise, the thermal abundance of dark matter would over-close the universe (if freeze-out occurs when the dark matter is relativistic) or cause unacceptably large changes to the neutrino and/or photon energy densities (non-relativistic freeze-out).  Given that current measurements detect a thermal cosmic neutrino background at high significance~\cite{Ade:2015xua,Follin:2015hya,Baumann:2015rya}, observations require that a sub-MeV dark matter particle should be more weakly coupled to the Standard Model than neutrinos (at MeV energies).

In this paper, we will explore the space of viable models for sub-MeV dark matter and the implications for direct detection experiments.  We will consider two classes of direct detection signatures: scattering and absorption.  The largest possible scattering cross-sections for light dark matter require the presence of a light mediator.  Introducing a new light particle coupled to the Standard Model is highly constrained by lab-based, astrophysical and cosmological measurements.  Astrophysical constraints are particularly stringent for couplings to electrons and photons.  The couplings to nuclei are less severely constrained by astrophysics and can give rise to larger elastic cross-sections than with the electrons.  When mediator couplings to nuclei are large enough to bring the dark matter cross-section well-above the neutrino floor, the mediator was necessarily in thermal equilibrium prior to the QCD phase transition.  In this region, the mediator must decay to dark radiation to avoid over-closing the universe, thereby producing an increase in $\Neff$ of at least $\Delta \Neff = 0.09$.  This contribution will be tested with both Stage III and IV CMB experiments~\cite{Abazajian:2016yjj} and excluding this abundance of dark radiation would push the cross-section close to the neutrino floor, despite the weakness of other constraints in this parameter range.

Dark matter absorption is a more accessible experimental signature for light dark matter.  Furthermore, for sub-MeV masses, kinematics forbids dark matter decay to the charged fermions of the Standard Model.  While kinematics allows decays to photons or neutrinos, these decays can be highly suppressed if that dark matter is protected by a non-linearly realized non-Abelian symmetry.  It is natural to embed this symmetry in the flavor symmetry of the Standard Model to allow for preferred coupling to specific Standard Model fermions.  These couplings allow for dark matter absorption by either electrons or nuclei.  However, the dark matter is necessarily of non-thermal origin and the couplings to the Standard Model must be sufficiently small to avoid thermalization.  While these cosmological constraints are significant, they do not push the absorption cross-sections below the region that can be experimentally accessible.

This paper is organized as follows: In Section~\ref{sec:freezeout}, we review the observation that thermal sub-MeV dark matter is excluded by current observations.  In Section~\ref{sec:dd}, we discuss the class of models that would be observable through scattering with electrons and nuclei.  We show that these cross-sections are already highly constrained by a combination of lab, astrophysical and cosmological bounds.  Upcoming cosmological observations will probe the regions of currently allowed parameter space with the most experimentally accessible cross-sections with nuclei.  In Section~\ref{sec:natural}, we discuss models with dark matter absorption by electrons and nuclei and explain the strong limits from cosmology.

The paper is supplemented by three appendices: Appendix~\ref{app:Neff} computes the change to $\Neff$ for a thermalized mediator under various circumstances.  Appendix~\ref{app:abundance} discusses the origin of dark matter and the implications for the elastic cross-section.  Appendix~\ref{app:nobullet} discusses the bounds on the cross-sections for sub-components of the dark matter that avoid the bullet cluster constraints on dark matter self-interactions.

\section{Requirements for a Thermal Abundance}
\label{sec:freezeout}

In this section, we will review the predictions of thermal dark matter with $\md \ll 10$ MeV and how it is excluded by current observation~\cite{Boehm:2012gr,Ho:2012ug,Steigman:2013yua,Boehm:2013jpa,Nollett:2013pwa,Steigman:2014uqa,Nollett:2014lwa}.  We will focus on the qualitative features of $\md <1$ MeV rather than explaining the precise value of the current limits.  This qualitative understanding will be useful for deriving new bounds in later sections.

The most basic requirement for a model of dark matter is that it reproduces the observed abundance $\Omega_{c}h^2 = 0.12$.  For a dark matter particle of mass $\md$, the energy density is given by $\rho_{\rm dm} \approx \md n_{\rm dm}$, where $n_{\rm dm}$ is the number density.  To be compatible with observations, therefore
\beq
n_{\rm dm}= \frac{1}{\md} \, \Omega_c \rho_{\rm cr} \approx 1.2 \times 10^{-3} \, {\rm cm}^{-3} \times \left( \frac{1 \, {\rm MeV}}{\md}\right)
\eeq
For masses $\md > 10$ eV, the number density of dark matter particles is much smaller than the number of photons, $n_\gamma \approx 4\times 10^2 \, {\rm cm}^{-3}$.  Regardless of how the dark matter was produced, the final population was not simply determined by number density in thermal equilibrium at $T \gg \md$, when $n_{\rm dm} \simeq n_\gamma$.

For $\md >  10$ MeV, the necessary suppression is easily achieved through freeze-out at a temperature $T_{\rm F} < \md$.  When the temperature falls below the mass of the particle, the Boltzmann suppression of the number density in equilibrium can naturally explain the observed dark matter abundance.  However, as we will now review, when $\md \ll 10$ MeV, thermal equilibrium with the Standard Model (at $T \sim \md$) disrupts the freeze-out of neutrinos and the generation of primordial abundances of nuclei during BBN.  In essence, cosmological evolution for $T < 10$ MeV does not leave room for even one additional degree of freedom.

\begin{figure}[h!]
\begin{center}
\includegraphics[width=0.65\textwidth]{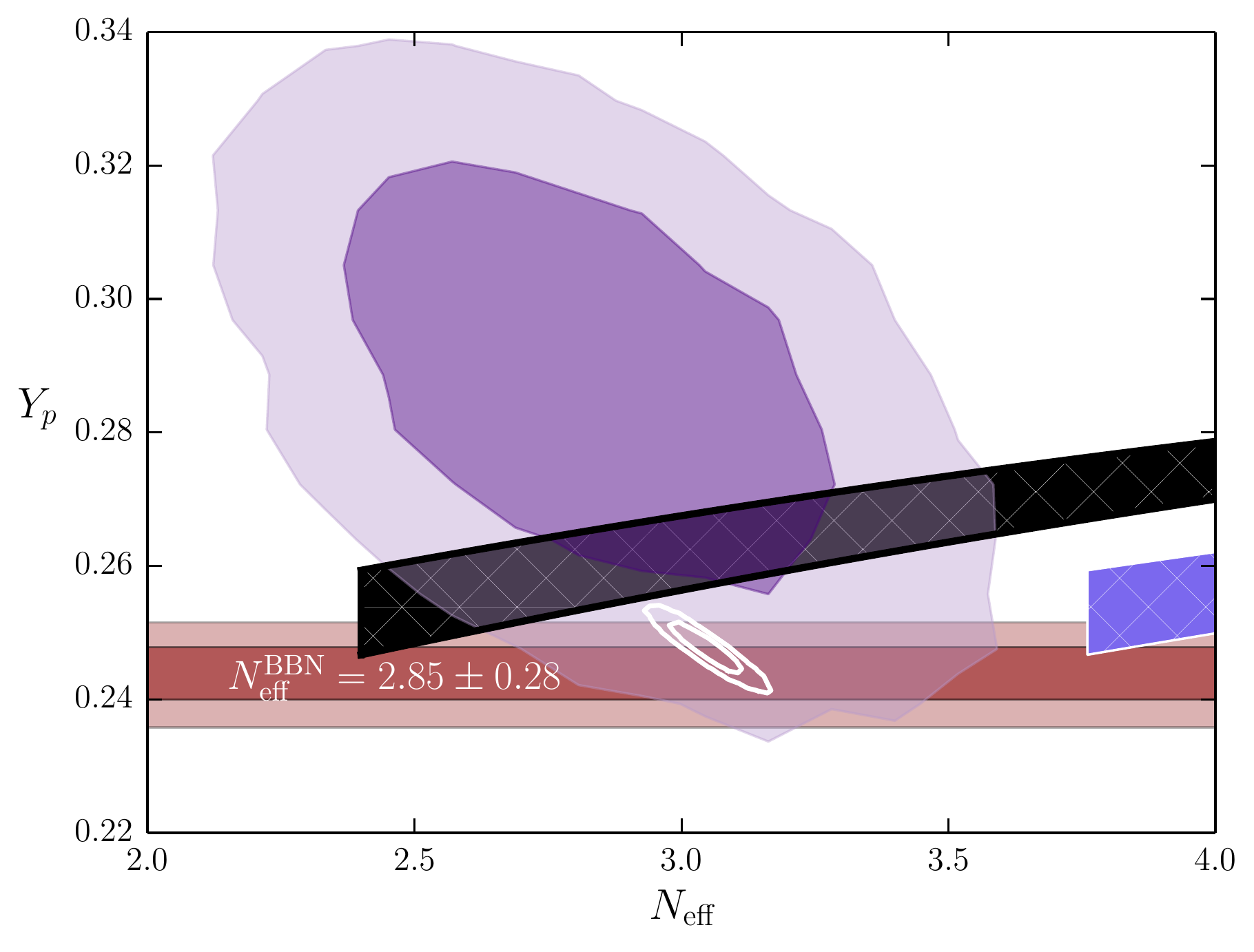}
\caption{Predictions for $\Neff$-$Y_p$ for thermal equilibrium with photons (black) and neutrinos (blue) assuming $g_{\rm DM} =1$ and $\md < 1$ MeV (fixing $\omega_b$ with a Planck prior). The length of the bands along the $\Neff$-axis allows for additional dark radiation to be added, which for photons can compensate the dilution of neutrinos from dark matter annihilation. The width of these bands along the $Y_p$-axis allows for the dark matter to contribute (or not) to $\Neff$ during BBN depending on its mass, with the largest value $Y_p$ corresponding to $\md \ll 100$ keV.  Current Planck 1 and 2$\sigma$ contours are shown in indigo~\cite{Ade:2015xua}.  The red region shows the range of $Y_p$ that is predicted by BBN if we impose the current constraint $\Neff^{\rm BBN} = 2.85 \pm 0.28$~\cite{Cyburt:2015mya} from the measurement primordial abundance, including D/H.  The direct measurement of primordial $Y_p$-alone yields a weaker constraint of $Y_p = 0.2465\pm 0.0097$.  In this way, we see that none of the hatched regions are consistent with all the information available from the CMB and abundance measurements. The white contours show forecasts for CMB Stage IV~\cite{Baumann:2015rya} (see also~\cite{Abazajian:2016yjj,Green:2016cjr}).  We see that future CMB data could completely rule out the entire space without the need for abundance measurements.  Figure adapted from~\cite{Baumann:2015rya}.}
\label{fig:Neff_thermal}
\end{center}
\end{figure}

The fundamental difficulty with $\md < 1$ MeV is that neutrinos begin to decouple at $T = 1-10$ MeV, which is necessarily larger than the decoupling of the dark matter.  The dark matter must annihilate at temperatures below its mass and will  either dilute or enhance the neutrino abundance which is out of equilibrium.  At these energies, the dark matter may couple to (1) photons\footnote{Couplings to electrons may or may not be a viable option depending on the mass of the dark matter.  If $\md > m_e$ then the annihilation to electrons is effectively the same a annihilating to photons because the electrons and photons are in equilibrium.  If $\md \ll m_e$, then the dark matter cannot annihilate efficiently when $T \sim \md$ and will over-close the universe.}, (2) neutrinos or (3) both.  It is easy to see that all of these scenarios are in tension with current observations:
\begin{enumerate}
\item Dark matter annihilating to photons dilutes the number/energy density in neutrinos relative to the photons.  From the conservation of comoving entropy:
\bea
\left(\frac{T_\nu}{T_{\gamma}} \right)^3 = \frac{2}{2 + \frac{7}{2}+ g_{\rm DM}} \ .
\eea
The resulting change to energy density in neutrinos is expressed as
\beq
\Neff = \frac{8}{7} \left(\frac{11}{4}\right)^{4/3} \frac{\rho_\nu}{\rho_\gamma} = 3\times \left(\frac{1}{1+ \frac{2}{11} g_{\rm DM}} \right)^{4/3} \leq 2.4 \ .
\eeq
where we used $g_{\rm DM} \geq 1$ for the final inequality.  More details of the calculation of $\Neff$ in this and other scenarios can be found in Appendix~\ref{app:Neff}.

\item Thermal equilibrium with neutrinos would heat the neutrinos after being diluted by electrons such that 
\beq
\left(\frac{T_\nu}{T_{\gamma}} \right)^3 = \frac{4}{11} \frac{\frac{21}{4} + g_{DM}}{\frac{21}{4}}
\eeq
which means that 
\beq
\Neff = 3 \times \left(1+ \frac{4}{21} g_{\rm DM} \right)^{4/3} \geq 3.78 \ ,
\eeq
where again we used $g_{\rm DM} \geq 1$.
\item Finally, if everything is in equilibrium with the dark matter until after the electrons and positrons annihilate and then the dark matter annihilates in equilibrium, then $T_\nu = T_\gamma$ and
\beq
\Neff  =3\times  \left(\frac{11}{4}\right)^{4/3} \approx 11.5 \ .
\eeq
\end{enumerate}
These numbers are all excluded at more than 3$\sigma$ for the current limit $\Neff=3.04\pm 0.18$~\cite{Ade:2015xua}.  Therefore, the only loopholes either require further modifications of the model to cancel these effects or to dial the couplings such that non-equilibrium production of neutrinos and/or photons is just right (notice that there is no way to make it work in equilibrium). 

For the particular case of coupling to photons, one might conclude that we can easily evade the bounds by adding some additional form of radiation to compensate for the diluted contribution of the neutrinos.  However, one cannot add this radiation without introducing significant changes to BBN.  Because $\md < 1$ MeV, the dark matter will contribute significantly to the expansion rate when $T > \md$ and therefore alters primordial abundances.  Furthermore, when the dark matter does annihilate it will dilute not only the neutrinos but also any additional dark radiation as well.  As a result, in order to add $\Delta \Neff^{\rm CMB}$ to $\Neff$ at recombination to avoid CMB constraints, we must introduce $\Delta \Neff^{\rm BBN} \approx (1+ \tfrac{2}{11} g_{DM})^{4/3} \Delta \Neff^{\rm CMB}$ to $\Neff$ during BBN.  Adding dark radiation during BBN alters the Helium fraction, $Y_p$ because it changes the expansion after neutron decoupling as well as other primordial abundances such as Deuterium.   The predictions for such a scenario are shown in Figure~\ref{fig:Neff_thermal} and are excluded by at least 2$\sigma$ over the entire parameter space.

\vskip 10pt
\noindent {\it Summary:}  \hskip 2pt Dark Matter with a mass $\md < 1$ MeV cannot get its abundance from thermal equilibrium with the Standard Model at temperatures $T \sim \md$.  Under minimal assumptions, it is excluded by CMB measurements of $\Neff$ for equilibrium with photons, neutrinos or both photons and neutrinos.  Simple attempts to evade these constraints by adding extra sources of radiation are excluded by measurements of primordial abundances.  While this description is based on the qualitative features, precise limits on the mass have been derived for each of these scenarios in~\cite{Ho:2012ug,Steigman:2013yua,Boehm:2013jpa,Nollett:2013pwa,Steigman:2014uqa,Nollett:2014lwa} .  
\section{Direct Detection through Scattering}
\label{sec:dd}
In a conventional picture of thermal freeze-out, it is necessary that the dark matter is coupled to the Standard Model.  To achieve the observed abundance, these couplings are required to be large enough that they make an enticing target for direct detection experiments.  On the other hand, for sub-MeV dark matter, one must exclude couplings large enough to bring the dark matter into thermal equilibrium. These upper-limits on the couplings of dark matter to the Standard Model leave uncertain the experimental prospects for viable models of sub-MeV dark matter.  In this section, we will explore these constraints and how they impact the experimental scattering cross-sections of dark matter with electron or nuclear targets.  

We will discuss these constraints within the context of a simplified model~\cite{Alves:2011wf} for the dark matter, involving a scalar mediator.  While the detailed constraints are specific to this model, any simplified model can be similarly constrained by these considerations.  In particular, the requirement that the dark matter was never thermalized will have significant implications for the coupling to the standard model, in much the same way that stellar cooling places limits the coupling to additional light fields.  Furthermore, our choice of models is intended to produce the largest possible experimental signature.  There are plenty of examples of light dark matter candidates that evade our bounds but would also evade direct detection (e.g. sterile neutrinos).  

We will defer discussion of the origin of dark matter to Appendix~\ref{app:abundance} and evaluate constraints on the elastic cross-section between dark matter $\chi$ and the Standard Model assuming that the correct abundance is generated but not necessarily by a thermal mechanism.  Elastic scattering of light dark matter will deposit tiny amounts of energy in a detector and thus inelastic processes might be preferable for direct detection experiments. But, the inelastic scattering cross-section can be expressed as a product of the elastic scattering cross-section and an inelastic form factor, typically suppressing the inelastic cross-section. Thus, our limits on the elastic scattering cross-section should be viewed as an upper bound, satisfying all cosmological, astrophysical and laboratory constraints.

\subsection{Simplified Model}
\label{subsec:simmodel}

Large experimental cross-sections require mediators that are light enough to be produced at relatively low temperatures in the early universe and/or in stars (even if they can be integrated-out for the purpose of direct detection).  To discuss cosmological/astrophysical constraints, our model must include both the dark matter ($\chi$) and mediator ($\phi$) particles. Our results can be summarized in terms of the model 
\begin{equation}\label{eq:model1}
\mathcal{L} \supset \frac{1}{2} \md \,  \lambda  \phi \chi^2  +   g_{N} \phi N \bar{N} + g_{e} \phi E \bar{E} -\frac{1}{2} \md^2 \chi^2 -\frac{1}{2} m^2_{\phi} \phi^2 \ ,
\end{equation}
which describes the (real-scalar) dark matter $\chi$ scattering off either a nucleon ($N$, coupling $g_N$)  or electron ($E$, coupling $g_E$). 

The couplings $g_e$ and $g_N$ are constrained by a variety of bounds. This includes short distance force experiments,  collider bounds and stellar/cosmological constraints. The bounds on $g_N$ and $g_e$ are strong functions of the mass $m_{\phi}$ of the mediator. Short distance force experiments dominate for $m_{\phi} < 100 \text{ eV}$, stellar bounds are strong for $m_{\phi} < 10$ keV, cosmological constraints and collider bounds are important for heavier masses. The interaction cross-section between dark matter and a Standard Model fermion  can be written as
\begin{equation}\label{eq:cross1}
\sigma_{\chi f \rightarrow \chi f} = \frac{\lambda^2 g_f^2}{4 \pi} \frac{\md^2}{\left( \left(\md v\right)^2  + m_{\phi}^2\right)^2}
\end{equation}
where $g_f$ is either $g_e$ or $g_N$ depending upon the fermion (electron, nucleon) that scatters with the dark matter.  This cross-section is valid in the limit $\md \ll m_e, m_N$, which will be the case for most of the parameter space of interest.  Naively, it would seem that the largest direct detection cross-section would be obtained by taking $m_{\phi} \rightarrow 0$. But since the bounds on $g_f$ are strongly dependent on $m_{\phi}$, this is not the case. For elastic/inelastic scattering, in light of proposed experimental technologies\footnote{Note that the cosmological bound of $\md >10$ keV~\cite{Viel:2013apy} applies only to thermal relics, which is not the case here because of the constraints from Section~\ref{sec:freezeout}.}, our primary interest is for dark matter with mass $\geq$ keV. To obtain the maximum allowed cross-sections, we will thus concentrate on mediators with mass $\geq 1$ eV, below which the direct detection scattering cross-section does not change, while bounds on the long range force mediated by $\phi$ between Standard Model particles gets considerably stronger. 

Bounds on dark matter self-interaction constrains $\lambda$ independently of $g_f$. The self-interaction scattering cross-section is constrained to be $\lessapprox 10^{-25} \text{ cm}^2/\text{GeV}$ from observations of the bullet cluster~\cite{Markevitch:2003at}. This bound is stringent and further suppresses the direct detection cross-section. However, if $\chi$ is less than $\sim$ 10 percent of the dark matter abundance, this bound does not apply. The experimental and astrophysical limits on $g_f$ are of course independent of $\lambda$ and constrain the experimental cross-sections for even such a sub-component. In Appendix~\ref{app:nobullet}, we will  show the limits on the cross-section when the self-interaction bound is not applied, with the caveat that the larger cross-section in this case is only possible for a sub-component of dark matter. 


\subsection{Electron Interactions}
\label{subsec:electrons}
Cosmology and astrophysics are the dominant bounds on $g_e$ for light mediators with $m_{\phi} \ll $ 10 MeV. In the range MeV -- GeV, bounds are placed by beam dump experiments while colliders are relevant for masses above GeV.  For mediator masses in the range MeV - GeV, sub-MeV dark matter is also constrained by SN1987A. 

\begin{figure}[h!]
\begin{center}
\includegraphics[width=0.65\textwidth]{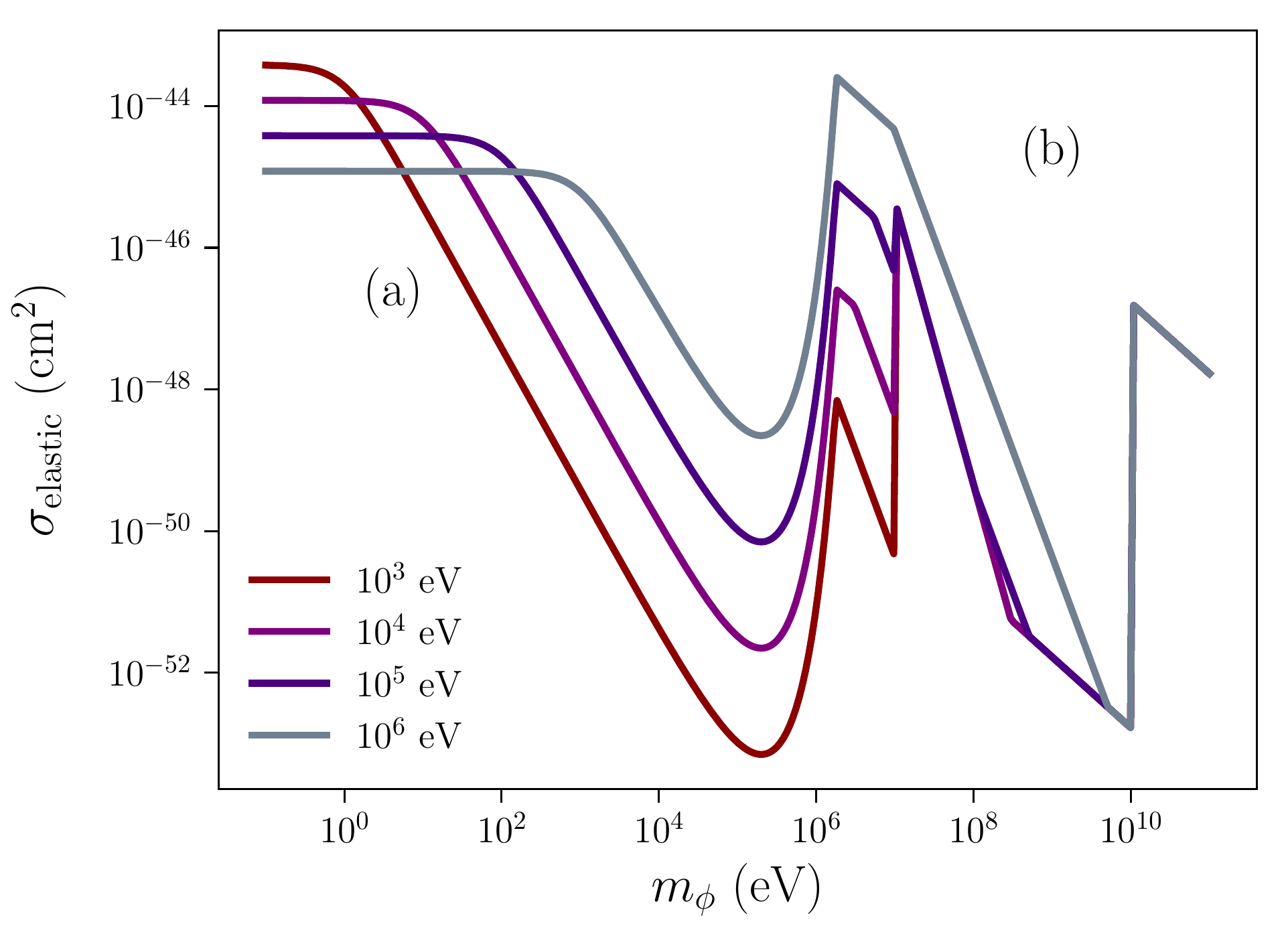}
\caption{Limits on the dark matter-electron elastic cross-section as a function of mediator mass for various dark matter masses in the range keV--MeV, as indicated in the legend.  The region (a) is primarily constrained by white dwarf cooling while region (b) is limited by B-factory searches and cosmology.  The sharp feature at $m_\phi \sim 10^{5-6}$ eV arises when $m_\phi$ approaches the white dwarf fermi-momentum, $k_f \approx 400$ keV and the additional cooling rate falls exponentially.   A second feature at $m_\phi  \approx 10$ MeV indicates where thermalization of the mediator becomes possible because it can decay to electrons or neutrinos before neutrino decoupling.  Neutrino decoupling is a gradual process that begins around 10 MeV and therefore the precise localization of this feature requires more care.}
\label{fig:electron_cross}
\end{center}
\end{figure}

If we focus on the region $m_{\phi} <$ 10 MeV, the dominant bounds on $g_e$ arise from cosmology and astrophysics.  Specifically, this region is dominated by the two constraints (with the relevant region for Figure~\ref{fig:electron_cross}):
\begin{itemize}
\item White dwarf cooling constrains $g_e < 8.4\times10^{-14}$~\cite{Hansen:2015lqa} when the mediator is light enough to be produced from $ e +e\to e+e+\phi$ using $k_f \approx 400$ keV. This constraint is dominant in {\bf region (a)}.
\item A thermal abundance of $\phi$ from thermal equilibrium with electrons and positrons would contribute $\Delta \Neff = 2.2$ and is easily excluded with current data.  Avoiding thermal equilibrium with electrons sets a limit $g_e < 2 \times 10^{-10}$~\cite{Baumann:2016wac}.  For $m_\phi> 10$ MeV, the mediator could decay before neutrino decoupling but cannot bring the dark matter into equilibrium (see Appendix~\ref{app:abundance}), which constrains either $\lambda$ or $g_e$.  These constraints are dominant in {\bf region (b)}.
\end{itemize}
For $m_{\phi} >$ 10 MeV, the thermalization of $\phi$ does not disrupt neutrino freeze-out or BBN, but we can impose the collider constraint
\begin{itemize}
\item B-factory searches for direct axion production limit~\cite{Izaguirre:2016dfi} $g_e < 10^{-3}$ for $m_{\phi} <$ 10 GeV .  This constraint is dominant in {\bf region (b)}.
\end{itemize}
Finally, we have the constraint on $\lambda$ from the bullet cluster~\cite{Markevitch:2003at}
\beq\label{eq:bullet}
\sigma_{\chi} = \frac{1}{8 \pi}\lambda^4 \frac{\md^2}{(m_{\phi}^2 + v^2 m_{\rm dm}^2)^2} < 1 \, {\rm cm}^2 \, / \, {\rm g} \times \md \ .  
\eeq
From these four constraints, we find the largest cross-section possible as a function of $m_\phi$ and $\md$ as shown in Figure~\ref{fig:electron_cross}.
 
 The qualitative feature of the limits on the electron-dark matter cross-section is that stellar constraints dominate for lower masses, $m_\phi < 400$ keV.  A variety of constraints from stellar cooling exist at this level but the strongest limits on the electron coupling arise in white dwarfs.  This is also important that due to the large density of electrons, the Fermi-momentum of $400$ keV is much higher than the thermal momentum of 1 keV.  As a result, the electrons are effectively relativistic.  This is not true of nucleons which is largely responsible for the qualitative differences between bounds on the electron and nucleon couplings.  Including the constraint from the bullet cluster, when $v \, \md < m_\phi < 400$ keV,  the cross section limit scales as $\sigma \propto m_\phi^{-2}$.
 
As the mediator mass approaches the Fermi-momentum, the mediator-induced cooling becomes exponentially suppressed and quickly weakens the bounds until we reach the limit set by cosmological constraints.  When $m_e < m_\phi < 10$ MeV,  forbidding equilibrium at $T = m_\phi$ (rather than $T=m_e$) requires that $g_e < 2 \times 10^{-10} \,  (m_{\phi} / m_e)^{1/2}$.  Naively the cosmological limits do not apply above 10 MeV since the mediator would annihilate before neutrino decoupling.  However, as described in Appendix~\ref{app:abundance}, if $\lambda$ is sufficiently large to bring $\chi$ into equilibrium then the entropy carried by the dark matter will produce a change to $\Neff$ that is already excluded by observations.   As a result, Figure~\ref{fig:electron_cross} shows the cross-section above 10 MeV is still strongly contained because of this additional bound. 

For mediators heavier than $\sim$ GeV, one could try to avoid these cosmological bounds by reheating the universe below 100 MeV. In this case, for sub-MeV dark matter, there are two important bounds. First, the dark matter would be emitted in SN1987A. For mediators of mass $\gtrsim$ GeV, at SN1987A temperatures $\sim 20$ MeV, the interaction between electrons and the dark matter is the same higher-dimension operator as observed in direct detection experiments. If the produced dark matter escapes the supernova, it can lead to enhanced cooling. The elastic cross-section necessary to avoid this bound would be $\lessapprox 10^{-50} \text{ cm}^2$, well beyond the scope of direct detection. However, if the dark matter is more strongly coupled, it can thermalize within the supernova. In this case, there might be an additional bound from requiring that the produced dark matter does not cause too many events in Super Kamiokande or contribute to a diffuse background of hot dark matter particles (much like the diffuse supernova neutrino background) that may have lead to events in experiments such as XENON. Moreover, this region is also constrained by LEP bounds on light dark matter \cite{Fox:2011fx}.  These bounds jointly squeeze the available parameter space in this window for sub-MeV dark matter. A more detailed analysis\footnote {P.W. Graham, S. Rajendran and G. M. Tavares, in progress} is necessary to figure out if there is a sliver of parameter space that is still allowed by these bounds.

We note that these constraints also rule out exotic electron - neutrino interactions that have been proposed to explain DAMA  \cite{Harnik:2012ni}. These models typically require the coupling of a massive vector boson ({\it e.g.} B-L) with gauge couplings $g \gtrapprox 10^{-8}$ and mass less than MeV. These vector bosons will be in thermal equilibrium with the Standard Model at temperatures $\sim$ MeV, after neutrino decoupling. As we have seen, the entropy in these bosons cannot be easily removed. Moreover, for B-L  bosons with mass less than $\sim$ MeV and gauge couplings $\gtrapprox 10^{-6}$, electrons and neutrinos would be coupled more strongly than the weak interactions at MeV temperatures, changing the neutrino temperature at decoupling. This additional entropy would appear as dark radiation and is similarly constrained. 

The most stringent bounds in these scenarios apply directly to the mediator $\phi$ and not on the light dark matter particle $\chi$. Once $\phi$ satisfies cosmological and stellar constraints, the production of $\chi$ would be sufficiently suppressed to avoid these bounds. While this is typically true of any model of light mediators, the argument fails for one specific case - when the mediator is a hidden photon that is kinetically mixed with electromagnetism. In this case, the Lagrangian  (in the mass basis) can be expressed as 
\begin{equation}
\mathcal{L} \supset F_{\mu \nu}^2 + G_{\mu \nu}^2 + m^2 B_{\mu}^2 + g_{\chi} J^{\mu}_{\chi} B_{\mu} + e J^{\mu}_{e} \left(A_{\mu} + \epsilon B_{\mu}\right)
\end{equation}
where $F_{\mu \nu}$ is the gauge field strength of electromagnetism ($A_{\mu}$), $G_{\mu \nu}$ is the gauge field strength of the hidden photon $B_{\mu}$. In this case, as is well known, the effects of the mediator decouple with its mass $m$ and thus bounds on the mediator are weakened in the limit $m\rightarrow 0$. However, the dark matter $\chi$ does not decouple when $m\rightarrow 0$ - it is in fact charged under electromagnetism with a charge $\sim \epsilon g_{\chi}$. In processes where the momentum transfer $q \gg m$, $\chi$ behaves as a milli-charged particle and would thus be constrained by cosmology and stellar bounds. These bounds imply that the effective charge $g_{\chi} \epsilon \lessapprox 10^{-13}$ for $m_{\chi} \lessapprox 10$ keV (stellar bounds) and $g_{\chi} \epsilon \lessapprox 10^{-11}$ for $\md \gtrapprox 10$ keV (cosmology), leading to cross-sections that are at least as small as $10^{-42} \text{ cm}^2$ ($m_{\chi} \gtrapprox$ 10 keV) and  $10^{-45} \text{ cm}^2$ ($\md \lessapprox$ 10 keV) (the long range coulombic enhancement has been cut-off at the angstrom scale, beyond which electric fields are screened in high density matter), without including additional inelastic or phase-space suppressions. 

In the models considered above, the dark matter was a scalar. Similar bounds can be obtained for the fermionic case. However, if the dark matter was a Dirac fermion, there are two additional couplings that are of interest --- namely, the dark matter may carry electric or magnetic dipole moments under electromagnetism. The light mediator in this case is the photon and it can give rise to enhanced scattering.  Much like axions, the scale $\Lambda$ suppressing the dipole operator is constrained by astrophysical bounds. For dark matter masses less than $400$ keV, limits from the cooling of white dwarfs require $\Lambda \gtrapprox 10^{10}$ GeV.  For  dark matter in the mass range 400 keV -- 20 MeV, there are constraints from energy loss in SN1987A: the range $10^{6}$ GeV $ \lessapprox \Lambda  \lessapprox 10^{9}$ GeV is excluded. For $\Lambda \gtrapprox 10^{6}$ GeV, the dark matter is thermalized within the supernova and does not lead to additional cooling. In this case, much like the hadronic axion window, there is an additional constraint from ensuring that the dark matter produced in SN1987A does not cause too many events in Super Kamiokande. Since the dark matter interacts through a dipole operator, it can scatter off electrons in Super Kamiokande,  requiring $\Lambda \lessapprox 10^{5}$ GeV. However, in this window, the dark matter is thermalized at MeV temperatures  and is thus ruled out (see e.g.~\cite{Sigurdson:2004zp,Dvorkin:2013cea} for further discussion of the limits from cosmology). Thus, we take the bound on $\Lambda$ to be $\gtrapprox 10^{10}$ GeV for dark matter masses $\lessapprox$ 400 keV and $\Lambda \gtrapprox 10^{9}$ GeV in the range 400 keV -- 20 MeV . These bounds apply to both magnetic and electric dipole operators since these processes occur at high energy. In direct detection, these operators yield different scattering cross-sections since the electric dipole scattering is enhanced by the long range electric force of electrons and nuclei. The electric dipole scattering cross-section is 
\beq
\sigma \approx  \frac{2 \alpha}{ \Lambda^2} \frac{\left(m_{\text{dm}} \right)^4 v^2}{\left( \left(m_{\text{dm}} v\right)^2 + \left(2 \text{ keV}\right)^2  \right)^2} \ ,
\eeq
while the magnetic dipole scattering cross-section is $\sigma \approx 3 \alpha / \Lambda^2$. We observe that the electric dipole scattering cross-section is enhanced at low velocity as expected from the long range electric field of the electron. However, this enhancement is cut off for momentum transfers less than 2 keV ($\sim$ angstrom) where electric fields in high density matter are screened. It can be seen that the electric dipole operator yields scattering cross-sections comparable to the case of light mediators considered earlier. 

\subsection{Nucleon Couplings and Dark Radiation}\label{subsec:nucleon}

The bounds on the couplings to nucleons are more complicated than electrons for several reasons.  First, the cosmological nucleon abundance is exponentially suppressed for $T \ll 1$ GeV.  Second, current data allows a light relic to be in equilibrium before the QCD phase transition but could be excluded with future measurements as we will explain below.  Finally, astrophysical constraints on nuclear couplings are less stringent due to the small velocities of the nucleons in stars and due to trapping effects in supernovae.  

\begin{figure}[h!]
\begin{center}
\includegraphics[width=0.49\textwidth]{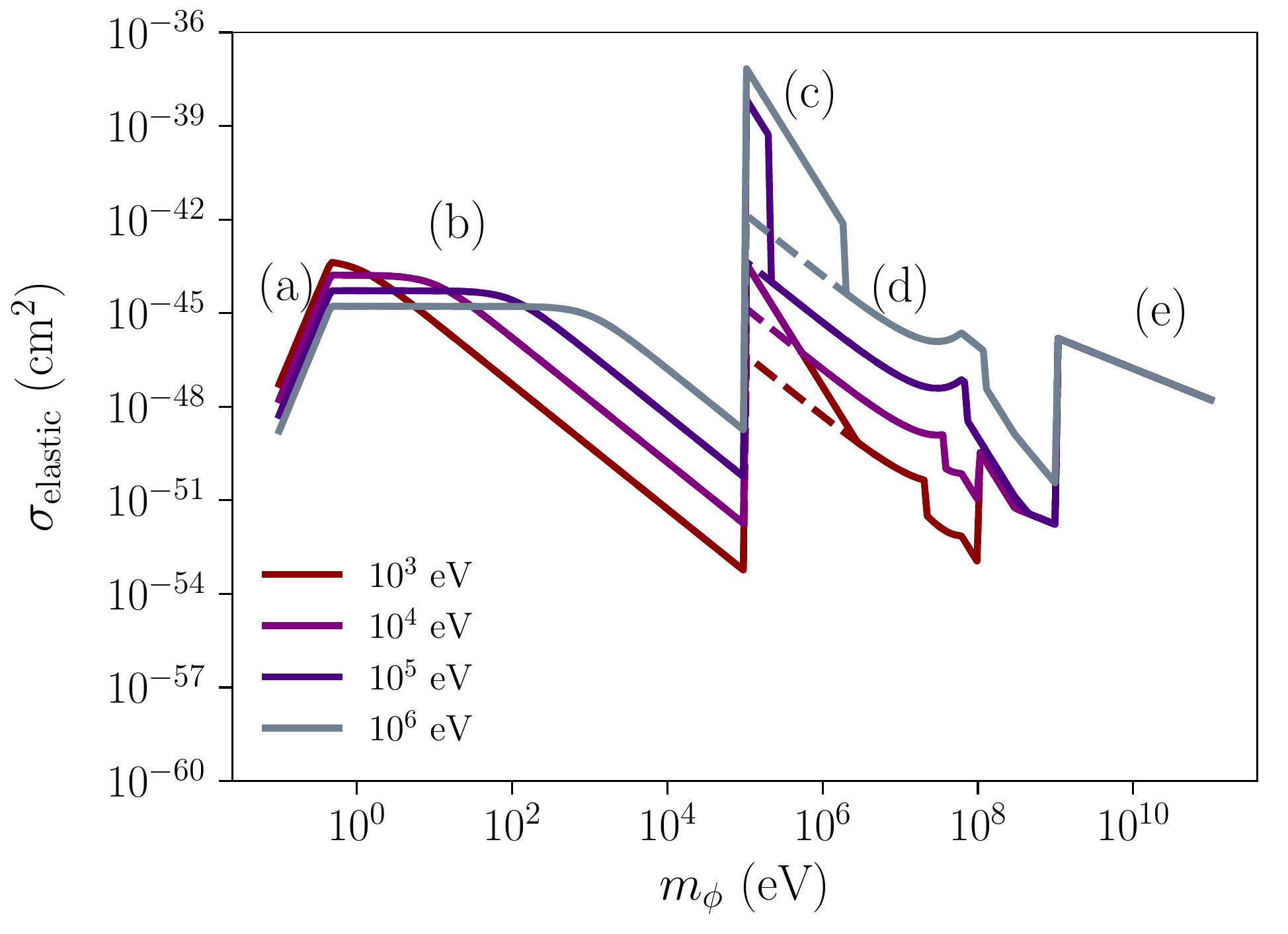}
\includegraphics[width=0.49\textwidth]{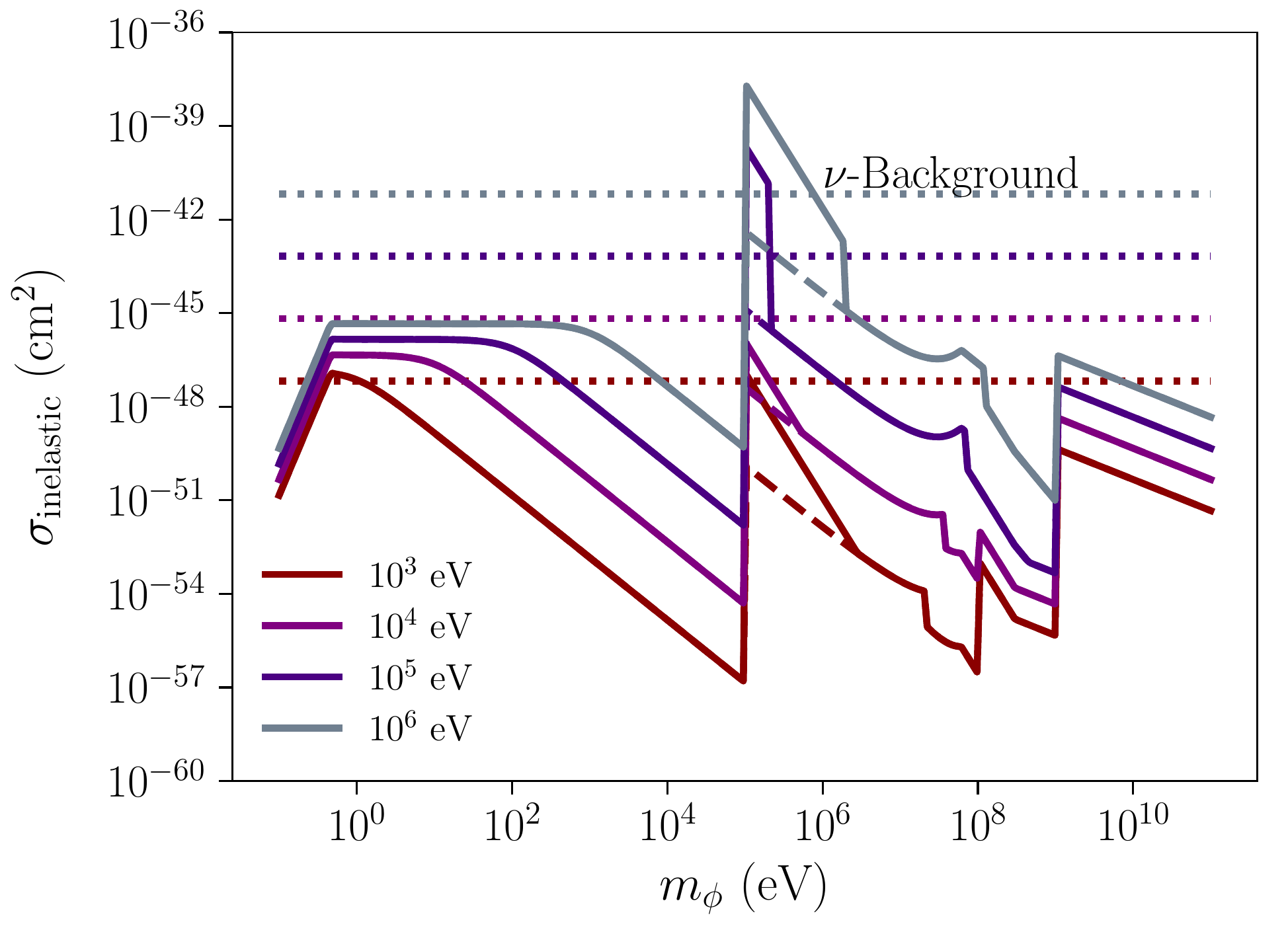}
\caption{{\it Left:}  Limits on the dark matter-nucleon elastic cross-section as a function of mediator mass for various dark matter masses.  The dashed (solid) lines show the cross-section with (without) a future cosmological constraint that excludes $\Delta \Neff =0.09$. The regions indicated by labels (a-e) are primarily constrained by (a) long range forces (b) cooling of HB stars (c) SN1987A (d) thermalization of mediator (because of mediator decay to dark matter) (e) meson decays, mediator decay to dark matter (limit on $\lambda$).
If a constraint on $\Neff$ forbids thermalization of the mediator above the QCD phase transition, then the trapping window allowed by SN1987A is excluded and the dashed line shows the region (c) is eliminated.  {\it Right:}  Limits on the dark matter-nucleon inelastic cross-section as a function of mediator mass for various dark matter masses, using $A=130$.  The dashed (solid) lines show the cross-section with (without) the constraint from excluding $\Delta \Neff =0.09$ in the CMB.  The dotted lines show the neutrino background expected for the same momentum transfer as a dark matter part of the appropriate mass.  Note that a future measurement excluding $\Delta \Neff =0.09$ would exclude a large fraction of the viable parameter space above the associated neutrino background.}
\label{fig:nucleon_cross}
\end{center}
\end{figure}

There are also a broader range of constraints on nucleons, compared to electrons.  The mediator coupling to nucleons, $g_N$, is most limited by (including the relevant region in Figure~\ref{fig:nucleon_cross})
\begin{itemize} 
\item Long range force experiments~\cite{Adelberger:2003zx} require that $g_N \lessapprox 10^{-12} (m_\phi / {\rm eV})^3$.  For our purposes we note that the constraints extend up to $m_\phi < 100$ eV. This constraint is dominant in {\bf region~(a)}.
\item For mediator masses below $\sim 100$ keV, bounds from the cooling of horizontal branch (HB) stars sets $g_N \lessapprox 10^{-13}$. The dominant production mechanism of these particles is through non-relativistic bremsstrahlung of the mediator during the coulomb scattering of nuclei. This constraint is dominant in {\bf region~(b)}. 
\item SN1987A excludes $3\times 10^{-10} < g_N < 3\times 10^{-7}$~\cite{Raffelt:1999tx} for $m_\phi < 30$ MeV.   This constraint is dominant in {\bf region~(c)}. The region $g_N > 3\times10^{-7}$ is allowed because the mediators become trapped.  This constraint is dominant in {\bf region~(d)}.
\item Meson decays require that $g_N < 10^{-6}$ for $m_\phi < m_\mu$ and $g_N < 10^{-3}$ for $m_\phi < 1$ GeV~\cite{Essig:2010gu}. This constraint is dominant in {\bf region~(e)}.
\end{itemize}
In addition, we have the following constraints on $\lambda$ and $g_N$ from dark matter phenomenology:
\begin{itemize}
\item The bullet cluster limit as written in Equation~\ref{eq:bullet}.  This constraint is dominant in {\bf regions~(a-c)}.
\item Thermalization\footnote{Significantly stronger bounds can be derived by demanding that $\phi$ and $\chi$ were in equilibrium, including at temperatures $T\approx \md$ in particular.  In such a scenario, achieving the correct relic abundance for the dark matter cannot be achieved independently of the coupling to $\chi$ and is highly restricted.  Furthermore, the dark matter is necessarily of thermal origin and the bounds on warm dark matter~\cite{Viel:2013apy} apply.  We will not impose such a strong constraint on equilibrium as it is depends sensitively on additional details of the model and whether the correct abundance can be achieved while satisfying other constraints.  Imposing a stronger thermalization constraint would not qualitatively change the results Figure~\ref{fig:nucleon_cross} and has no effect on the dashed curves where the additional constraint on $\Neff < 0.09$ is imposed.}  of the dark matter at $T\gtrsim 300$ MeV would either over-close the universe and/or would lead to unacceptably large contribution to $\Neff$.  When $m_\phi < 2 \md$, forbidding thermalization requires $\lambda^4 < 8 \pi m_{\rm dm}^{-2} \, (H(T) \, T)|_{T\sim 300 \, {\rm MeV}} $ from $\phi \phi \to \chi \chi$.  When $m_\phi \geq 2 \md$, mediator decay can thermalize the dark matter directly and the thermalization constraint becomes $\lambda^2 < 8 \pi m_{\rm dm}^{-2} \, (H(T) \, T)|_{T\sim {\rm max}(m_\phi, 300 \, {\rm MeV})}$. This constraint is dominant in {\bf region (d) and (e)}.
\end{itemize}

It is worth noting that in any given part of parameter space, there is only one dominant bound on each of the couplings.  These constraints\footnote{Although the constraints on individual couplings have a proper statistical meaning (typically the 2$\sigma$ or 95\% C.I.), the ``bounds" on the cross-section we derive from them do not have a precise statistical interpretation.  Given the range of scales involved, our qualitative bounds are sufficient for our conclusions.} can be translated into an elastic cross-section for scattering with a proton or neutron, $\sigma_{\rm elastic}$, using Equation~(\ref{eq:cross1})\footnote{For light mediators, the elastic scattering cross-section can be enhanced at low momentum transfer. However, at low momentum the energy deposited in the collision is suppressed by $q^2$. There are no known experimental methods to measure these ultra-low energy recoils.}.  However, since the energy deposited in an elastic collision is very small, it is likely that these dark matter candidates will be probed by inelastic processes. In particular, one may consider inelastic collisions that excite molecular bonds in a gas. This would provide a useful upper-bound for inelastic collisions - if the detector atoms were in a higher density fluid or solid lattice, the corresponding phonon excitations will have similar inelastic form factors while being subject to additional kinematic constraints that may further suppress the rate. The inelastic cross-section to excite a molecular state consisting of atoms with atomic mass $A$ is: 
\beq
\sigma_{\rm inelastic} = \sigma_{\rm elastic}  \times  \left( \frac{A^2 q^2}{A \, m_{\rm nucleon} \,  \omega(q) }  \right) \ ,
\eeq
where $q = v m_{\rm dm}$ is the momentum transfer,  $\omega(q) = \frac{1}{2} \md v^2$ is the associated kinetic energy and $A m_{\rm nucleon}$ is the atomic mass of the target nucleus.  This cross-section can be compared to the event rate due to solar neutrinos~\cite{Gutlein:2010tq} elastically scattering off nuclei and depositing energies below eV. This rate is roughly $\Gamma_\nu \sim 10^7 \,  {\rm keV}^{-1}\, {\rm ton}^{-1} \, {\rm year}^{-1} $ in a detector using a heavy atom like Xenon. 

The implications of these constraints for direct detection experiments are shown in Figure~\ref{fig:nucleon_cross}.  For $m_\phi < 1$ eV, the long range force constraints are stronger than stellar constraints and the resulting cross-section lies well below the neutrino floor.  The long range force constraints decrease as $m_\phi^{-3}$ and becomes less important than stellar cooling for $m_\phi > 1$ eV.  For $100\, {\rm keV}< m_\phi < {\cal O}(1)$ MeV, the mediator is sufficiently heavy to evade constraints from HB stars and can still be consistent with the trapping region of the SN1987A constraint.  This is the one large region in the space of mediator and dark matter masses where the cross-sections lie well above the neutrino floor.

\begin{figure}[h!]
\begin{center}
\includegraphics[width=0.65\textwidth]{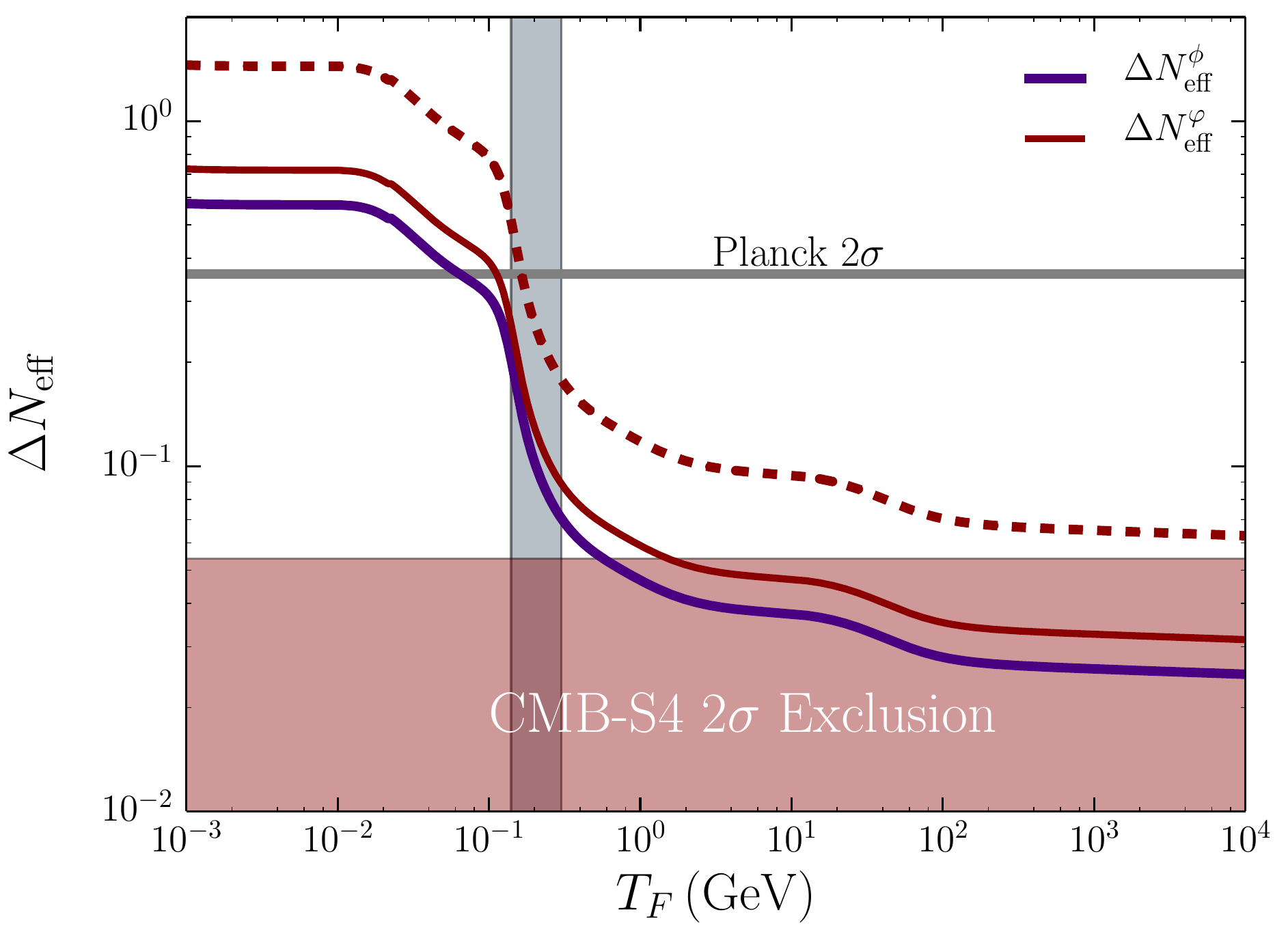}
\caption{The contributions to $\Neff$ from $\phi$ and $\varphi$ as a function of the temperature where $\phi$ decouples from the Standard Model, $T_F$, assuming $g_\varphi =g_\phi=1$.  The grey region indicated the temperature of the QCD phase transition.  The indigo line shows the contribution of $\phi$ to $\Neff$ when $m_\phi < 1$ eV and is stable.  The red line shows the contribution from $\varphi$ to $\Neff$ when it comes into equilibrium with $\phi$.  The solid (dashed) curve assumes $\varphi$ is not (is) in equilibrium at $T_F$.   We show the current 2$\sigma$ exclusion from Planck in grey and the projected 2$\sigma$ exclusion from CMB-S4 in red (assuming $\sigma(\Neff) =0.027$).  The case of interest for large $g_N$ has $T_F \approx 300$ MeV and would be detectable with CMB-S4 in all cases.  We used the lattice measurement of $s/T^3$ from~\cite{Borsanyi:2013bia} to determine $\Delta \Neff$ in the range $100 \, {\rm MeV} <T_F < 1 \, {\rm GeV}$. }
\label{fig:DNeff}
\end{center}
\end{figure}

When $g_N > 10^{-9}$, the mediator is thermalized before the QCD phase transition.  The limits on long range forces only permit couplings of this size when $m_\phi > 100$ eV.  This possibility is excluded either if $\phi$ is cosmologically stable (from over-closure) or if it decays to Standard Model particles (equivalent to equilibrium at $T< 10$ MeV).  Therefore, we must introduce a new light field $\varphi$ with $m_\varphi \ll 1$ eV into which $\phi$ can decay.  While there are, in principle, many ways to introduce new light degrees of freedom, they will all produce non-trivial contributions to $\Neff$, as discussed in Appendix~\ref{app:Neff}.  To avoid current limits, we should take $\varphi$ to be a real scalar such that $g_\varphi =1$.  The coupling to $\phi$ can be chosen such that $\varphi$ is out of equilibrium at the freeze-out temperature of $\phi$, $T_F \approx 300$ MeV, but is in equilibrium at $T=m_\phi$.  The resulting change to $\Neff$, $\Delta \Neff$, is given by 
\beq
\Delta \Neff = 2^{1/3} \times \Delta \Neff^\phi(T_F\approx 300 \, {\rm MeV} ) \approx 0.09 \ .
\eeq
The general behavior as a function of the freeze-out temperature $T_F$ is shown in Figure~\ref{fig:DNeff}.

Future CMB experiments like the Simons Observatory and CMB Stage IV~\cite{Abazajian:2016yjj} are expected to reach $\sigma(\Neff) \lesssim 0.045$ and $\sigma(\Neff) \lesssim 0.035$ respectively, and could exclude\footnote{Forecasts and limits for $\Lambda$CDM+$\Neff$ assume BBN consistency for $Y_p$, which means that we assume $\Neff$ at BBN and the recombination are the same.  However, when $1 \, {\rm eV} \ll m_\phi \ll 1$ MeV, this model predicts $\Delta \Neff^{\rm BBN} = 0.07$ and $\Delta \Neff^{\rm CMB} = 0.09$.  This difference is sufficiently small that it will not significantly reduce the sensitivity to our target.  See~\cite{Abazajian:2016yjj} for further discussion.  } these values of $\Delta \Neff$.  In the presence of an upper-limit $\Delta \Neff < 0.09$, we would be forced to add an additional constraint $g_N < 10^{-9}$ for $m_\phi < 10$ MeV, which would lead to $g_N < 3\times 10^{-10}$ from SN1987A\footnote{The mediator can also thermalize above the electroweak phase transition ($T \gg 100$ GeV) for some couplings small enough to avoid thermalization around the QCD phase transition~\cite{Baumann:2016wac}.  A limit of $\Delta \Neff < 0.034$ would exclude this possibility but the precise bound on $g_N$ depends on the reheat temperature.  Under reasonable assumptions, this bound is stronger than from freeze-out at the QCD phase transition and potentially SN1987A.}.  This measurement eliminates the sharp features in Figure~\ref{fig:nucleon_cross} and eliminates the allowed parameter space above the neutrino floor.  In this sense, upcoming CMB observations will test the remaining viable regions of the parameter space with large cross-sections. In the absence of signals in upcoming CMB observations, only one window in mediator mass would allow large $g_N$ -- if $m_\phi > 10$ MeV, it is possible to add couplings for $\phi$ to neutrinos (or electrons/photons if allowed by other bounds) that permit the mediator to decay to neutrinos (rather than $\varphi$) before neutrino decoupling and therefore could produce $\Delta \Neff=0$ even if the mediator was thermalized before the QCD phase transition.

Although thermalization of just the mediator is consistent with current observations, we cannot allow the dark matter to thermalize simultaneously.  Eliminating the entropy carried by the dark matter ultimately produces enough dark radiation to be nearly excluded by current observations, as discussed in Appendix~\ref{app:abundance}.  When $m_\phi <2 m_{\rm dm}$ this places a strong constraint on $\lambda$ from $\phi\phi  \to \chi \chi$.  When $m_\phi > 2 m_{\rm dm}$, thermalization can proceed easily through the decay $\phi \to \chi \chi$ and the resulting constraint is more stringent.  The easiest way to avoid this constraint is to forbid thermalization of the mediator in the first place, which requires $g_N < 10^{-9}$ which becomes $g_N< 3\times 10^{-10}$ from  SN1987A.  For this reason, we see that future constraint on $\Neff$ do not alter our sensitivity curves for $m_\phi > 2 m_{\rm dm}$ because thermalization has already been forbidden. 

Cosmological constraints played a central role in suppressing the cross-section when $m_\phi > 2 m_{\rm dm}$. For heavy mediators ($\gtrapprox$  GeV),  it is possible to evade these constraints by reheating the universe to temperatures below $100$ MeV. While such a low reheat temperature may not be aesthetically pleasing, it is allowed by data. When the cosmological bounds are removed, these mediators are  constrained by colliders. For $m_{\phi} \lessapprox$ GeV, rare meson decays limit $g_N \lessapprox 10^{-3}$~\cite{Essig:2010gu}. For heavier mediators, $g_N$ is constrained by Tevatron searches for mono-jets and missing energy~\cite{Bai:2010hh}. It can be checked that the allowed cross-sections for sub-MeV dark matter are below the neutrino floor for these cases. 

Finally, it is worth noting that the limits we have derived from cosmology typically require the the elastic cross-section to be significantly lower than the limits  found in~\cite{Dvorkin:2013cea} (or similar analyses for other interactions such as~\cite{Wilkinson:2013kia,Cyr-Racine:2013fsa,Wilkinson:2014ksa}).  Those limits were derived by introducing the cross-section as a parameter in the Boltzmann code used to compute the CMB anisotropies.  The key distinction is that in order to produce cross-sections at the limits of \cite{Dvorkin:2013cea}, the force must be mediated by a light particle that would be produced at relatively low temperatures during our thermal history. Mediators are physical particles which can produce significant observational implications that are not captured by the cross-section alone.  This is analogous to the need for a simplified model approach to dark matter searches at the LHC~\cite{Bai:2010hh,Fox:2011pm,Shoemaker:2011vi,Buchmueller:2013dya,Abdallah:2015ter}.  Of course, our bounds also include experimental and astrophysical bounds on the mediator and dark matter and are therefore not purely cosmological nor completely model-independent, unlike~\cite{Dvorkin:2013cea}.  

\section{Dark Matter Absorption}
\label{sec:natural}
In this section, we consider models where the dark matter can be absorbed by Standard Model particles, leading to the deposition of the rest mass of the dark matter into the detector rather than its kinetic energy alone. Unlike the models of elastic/inelastic scattering considered above, absorption implies that the dark matter lacks a conserved quantum number. Its stability is thus not guaranteed and requires additional structure. Of course, the lifetime of the dark matter can be made sufficiently long by suppressing its coupling to the Standard Model but at the cost of dramatically suppressing any direct detection signal.  It is interesting to ask if there might be theoretical structures that guarantee the stability of dark matter while permitting observable direct detection absorption cross-sections. 

A fruitful way to construct such models would be to take the mass of the dark matter to be small enough to kinematically forbid it from decaying to electrons. While additional structure is necessary to forbid decays to photons and neutrinos this kinematic suppression would, in principle, allow for large couplings to charged fermions. Hidden photons and axions are canonical examples of such models and their phenomenology (including stellar and cosmological bounds) is well studied \cite{Pospelov:2008jk}. These models are based on Abelian structures. Here, we discuss another class of models where the dark matter is the goldstone boson of a non-Abelian global symmetry. We argue that the non-Abelian structure provides extra protection for the Goldstone boson enabling it to have even larger couplings without causing it to decay. 

\subsection{Stability through Symmetry} 

To simplify the discussion, let us assume the dark matter is coupled to the Standard Model through
\beq\label{eq:darkint}
{\cal L}_{\rm int}  = \bar \psi_i  \left[e^{- i \mathbf{T} \chi/\Lambda_\psi} \, \mathbf{m} \, e^{i \mathbf{T} \chi / \Lambda_\psi}\right]^{ij} \psi_j  \ .
\eeq
where $\mathbf{T}$ is a generator of a subgroup of a $SU(3)$ flavor symmetry and $\mathbf{m}$ is the mass matrix.  By Taylor expanding the exponential, the leading interaction at low energies can include those we used for the mediator in the previous section\footnote{This coupling is very similar to the pseudo-scalar dark matter models of~\cite{Pospelov:2008jk} but we are explicitly using the non-abelian flavor symmetry to further suppress decays.}.  We have written the coupling in this form to indicate that $\chi$ is a pseudo-goldstone boson of a subgroup of $SU(3)$.  

The coupling to matter starts at linear order in $\chi$ which means that the dark matter is not guaranteed to be stable by a discrete symmetry.  Nevertheless, for $\md < 1$ MeV, kinematics forbid the dark matter from decaying to any charged fermions of the Standard Model and therefore this operator does not lead to decay directly.  In order to be a viable dark matter candidate, we must therefore avoid rapid decays to photons and neutrinos.  While this is a stringent constraint on an axion, $\chi$ is a pseudo-goldstone boson of a non-abelian subgroup of the flavor symmetry and the decays are more easily avoided.  Specifically, the non-abelian subgroups are non-anomalous and therefore the coupling $\chi F_{\mu \nu }\tilde F^{\mu\nu}$ is forbidden\footnote{We can also see this by noting that the transformation of $\chi$ includes terms that are nonlinear in $\chi$ and therefore the action would change by more than a total derivative.  One can see this more easily by including Goldstone bosons for all the generators of $SU(3)$ to simplify the analysis.  Giving these additional Goldstones masses above an MeV would allow them to decay to avoid any cosmological implications. }. This suppression of the photon coupling may also permit a natural way to incorporate the nucleo-phillic models discussed in section \ref{subsec:simmodel}.  Since neither the photons nor neutrinos are charged under the flavor symmetry, any coupling must take the form ${\cal O}_\chi \times {\cal O}_{A_\mu, \nu}$ where ${\cal O}_{\chi, A_\mu, \nu}$ are scalars under the flavor symmetry.  The leading decay we can write down is schematically
\bea
{\cal L}_{\rm decay} &=& \frac{1}{\Lambda_\gamma^2} {\rm Tr} [\boldsymbol{\lambda}^{\dagger } e^{-i \mathbf{T} \chi / \Lambda_\psi} \partial^2 e^{i \mathbf{T}  \chi / \Lambda_\psi} \boldsymbol{\lambda} ]  F_{\mu \nu} F^{\mu \nu} \nonumber \\
&\approx& |\lambda|^2 \frac{\md^2}{\Lambda_\gamma^2}  \frac{\chi}{\Lambda_\psi}  F_{\mu \nu} F^{\mu \nu}
\eea
where $\boldsymbol{\lambda}$ is the Yukawa couplings matrix and $\mathbf{T}$ is the broken generator of the flavor symmetry.  The decay rate is highly suppressed because of the symmetry, giving 
\bea
\Gamma &=& \frac{1}{64 \pi} \frac{\md^3}{\Lambda_\psi^2} \frac{\md^4}{\Lambda_\gamma^4} |\lambda|^4 \nonumber \\
&\approx& 8\times 10^{-30} \, {\rm s}^{-1} \, \left(\frac{\md}{1\, {\rm MeV}} \right)^6 \left(\frac{1 \, {\rm TeV}}{\Lambda} \right)^6 \left( \frac{|\lambda|^2}{10^{-6} } \right)^2 \ ,
\eea
where we used $\Lambda_\gamma = \Lambda_\psi  = \Lambda$.  LHC constraints require $\Lambda > 1$ TeV which implies a decay rate much longer than the age of the universe and consistent with $X$-ray/$\gamma$-ray~\cite{Essig:2013goa} and CMB limits~\cite{Slatyer:2016qyl} on decaying dark matter.

\subsection{Cosmological Constraints}

Just like the case of the light mediator, the coupling of dark matter to Standard Model fermions in Equation (\ref{eq:darkint}) allows for possible dark matter thermalization both at high temperatures (freeze-out) and at low temperatures (freeze-in).  However, unlike the mediator, any thermalization of $\chi$ is already excluded by current observations~\cite{Viel:2005qj} for $\md \gtrsim 10$ eV and would over-close the universe for $\md \gtrsim 200$ eV.  Therefore, we must impose the constraint that the dark matter was not thermalized in either regime.  The most conservative limits are derived from forbidding freeze-in because they apply to low reheat temperatures, $T_R \gtrsim 1$ GeV (except for the top quark coupling).  If $T_R > 1$ TeV, then we can derive a stronger constraint because the thermalization of the dark matter becomes more efficient with increasing temperatures.  The production rates are explicitly calculated in~\cite{Baumann:2016wac} and the resulting constraints on $\Lambda_\psi$ are shown in Table~\ref{tab:bounds}.  

The diagonal freeze-in constraints for quarks were not computed in~\cite{Baumann:2016wac} because of large thermal corrections that arise at temperatures $T\sim m_\psi$.  The constraint on the off-diagonal quark couplings were more easily computed because the leading effect does not depend on couplings to gluons, and is given by 
\beq\label{eqn:offdiag}
\Lambda_{ij} > 2.1 \times 10^9 \, {\rm GeV} \left(\frac{g_{\star,i}}{g_{\star,t} } \right)^{-1/4} \, \left(\frac{m_i}{m_t}\right)^{1/2} \ ,
\eeq
where $\psi_i$ is the heaviest fermion in the coupling and $g_{\star,i}$ is the number of effective degrees of freedom at $T = m_i$.  The diagonal coupling requires the emission of a gluon (or photon) and is suppressed by $\alpha_S^{1/2}$ but otherwise follows the same scaling as Equation~(\ref{eqn:offdiag}).  One can re-sum the soft thermal effects to determine a precise constraint following~\cite{Salvio:2013iaa}, but for our purposes the order of magnitude constraint that follows from (\ref{eqn:offdiag}) is sufficient.

\begin{table}[h!]
\begin{center}
 \begin{tabular}{c c c} 
 	Coupling  			 & Freeze-Out [GeV]	& Freeze-In [GeV]	\\
  \midrule[0.065em]
	$\Lambda_{ee}$						 & \num{6.0e7}		& \num{2.7e6}		\\[2pt] 
	$\Lambda_{\mu\mu}$	& \ \num{1.2e10}		& \num{3.4e7}		\\[2pt] 
	$\Lambda_{\tau\tau}$	 & \ \num{2.1e11}		& \num{9.5e7}		\\[2pt] 
	$\Lambda_{bb}$		& \ \num{9.5e11}		& --				\\[2pt]
	$\Lambda_{tt}$		& \ \num{3.5e13}		& --				\\
  \midrule
	$\Lambda_{\mu e}^V$	 & \num{6.2e9}		& \num{4.8e7}		\\[2pt] 
	$\Lambda_{\mu e}$	 & \num{6.2e9}		& \num{4.8e7}		\\[2pt] 
	$\Lambda_{\tau e}$	 & \ \num{1.0e11}		& \num{1.3e8}		\\[2pt] 
	$\Lambda_{\tau \mu}$	& \ \num{1.0e11}		& \num{1.3e8}		\\[2pt] 
	$\Lambda_{cu}^A$		& \ \num{1.3e11}		& \num{2.0e8}		\\[2pt] 
	$\Lambda_{bd}^A$		& \ \num{4.8e11}		& \num{3.7e8}		\\[2pt] 
	$\Lambda_{bs}$		 & \ \num{4.8e11}		& \num{3.7e8}		\\[2pt] 
	$\Lambda_{tu}$		 & \ \num{1.8e13}		& \num{2.1e9}		\\[2pt] 
	$\Lambda_{tc}$		& \ \num{1.8e13}		& \num{2.1e9}		\\[2pt] 
  \bottomrule 
 \end{tabular}
\caption{Constraints on couplings to matter where freeze-out assumes $T_{R} \approx 10^{10}$ GeV and freeze-in assume $T_{R} > m_f$.  This table is adapted from~\cite{Baumann:2016wac}. }
\label{tab:bounds}
\end{center}
\end{table}

After electrons, the only realistic targets for a direct detection experiment are nuclei, through the coupling to the $u$ or $d$ quarks.  The freeze-out limit assuming $T_R = 10^{10}$ GeV is $\Lambda_{uu} >5 \times 10^{8}$ GeV and $\Lambda_{dd} >1 \times 10^{9}$ GeV.  The best possible limit from freeze-in cannot be computed reliably because it depends on physics during the QCD phase transition.  From the scaling in Equation~\ref{eqn:offdiag}, one could place a bound $\Lambda_{uu,dd} > 10^6$ by forbidding freeze-in at temperatures above the QCD phase transition. 

Although these limits are minimal limits demanded by reheating above the QCD phase transition, more stringent conditions on $\Lambda$ are necessary in many circumstances.  Since the stability of dark matter follows from symmetry, the generators of $SU(3)$ will imply couplings to the 2nd and 3rd generation fermions with the same strength, $\Lambda$.  In Table~\ref{tab:bounds}, we show the bounds on these couplings adapted from~\cite{Baumann:2016wac} for freeze-in and freeze-out.  If we required, for example, that $\Lambda_{ee} = \Lambda_{\tau \tau}$ the constraints from freeze-out become four-orders of magnitude stronger (eight-orders of magnitude in the cross-section).

\subsection{Absorption Signatures}

In order to be consistent with BBN, we must assume that the universe  reheated to at least 10 MeV.  We therefore require that electrons cannot bring the dark matter into equilibrium below 10 MeV, which implies that $\Lambda_{ee} > 2.7 \times 10^6$ GeV.  For interactions with electrons, as shown in \cite{Pospelov:2008jk}, the absorption cross-section for $\chi$ is related to photo-absorption via: 
\beq
\sigma_{e\chi\to e \gamma} \approx  \frac{m_{dm}^2}{\Lambda_{ee}^2 \alpha}  \sigma_{e\gamma\to e \gamma} \ .
\eeq
The corresponding event rate for dark matter absorption is
\beq
R\approx 2 \times 10^{22} \, {\rm kg}^{-1} \, {\rm day}^{-1} \,  \frac{g_{ee}^2}{A} \, \frac{\md}{\rm MeV}  \, \left( \frac{\sigma_{e\gamma\to e \gamma}}{10^{-24} \, {\rm cm}^2} \right) \ ,
\eeq
where $g_{ee} \equiv 2 m_e / \Lambda_{ee}$.  Experimental limits on the absorption of light dark matter by electrons have been placed in~\cite{An:2014twa,Bloch:2016sjj} where they find limits of $g_{ee}  \lesssim 3\times \times 10^{-13}$ in the mass range 1 eV to 10 keV.  These experimental limits are stronger than our limit of $g_{ee} < 4\times 10^{-10}$  but are significantly weaker than white dwarf cooling bounds that apply in the same mass range.  Our cosmological bound is unique in that it is independent of mass and therefore cannot be avoided by kinematics or trapping effects as we increase the mass towards 1 MeV.  Furthermore, as discussed in the pervious section, cosmological limits also become competitive with white dwarf cooling when we require (by symmetry) a common coupling strength for third generation leptons as well.

The allowed absorption cross-sections can be significantly larger than the allowed scattering cross-sections we found in Section~\ref{sec:dd}.  For example, a given target with a photo-absorption cross-section of $\sigma_{e\gamma\to e \gamma} > 10^{-24} \, {\rm cm}^2$ (1 bn) and $\md \sim m_e$, our cosmological bound on $\Lambda_{ee}$ would allow a dark matter absorption cross-section of $\sigma_{e\chi\to e \gamma} > 10^{-41} \, {\rm cm}^2$ which is larger than any point in Figure~\ref{fig:electron_cross}.  This might seem surprising given that the cosmological limits on $\Lambda_e$ is identical to cosmological bounds on $g_e$ for the mediator in Section~\ref{subsec:electrons}.  The difference is that the scattering cross-sections are additionally suppressed by $\lambda$, which is constrained by the bullet cluster.  In this sense, our constraints on absorption are more analogous to the scattering constraints without the bullet cluster discussed in Appendix~\ref{app:nobullet}.  Of course, the operators that lead to absorption also produce elastic scattering but it is easy to see that the scattering cross-sections are suppressed by additional powers of $\Lambda_e$ and would be unobservable.

In addition to electron absorption, $\chi$ can also be absorbed by nuclei. Depending upon its mass, the signals could include excitation of phonons/vibrational levels of molecules, ionization of atoms or dislocations of atoms from crystal lattices.  Computation of these cross-sections is beyond the scope of this work.  Nevertheless, unlike the case of elastic or inelastic scattering, we do not expect to find a qualitative differences between the viable cross-sections for electrons and nuclei.  Specifically, the dark matter is stable and therefore thermalization would lead to over-closure.  Requiring that the dark matter was not thermalized also implies that it must satisfy the cooling constraint from SN1987a such that $\Lambda_p > 10^{9}$ GeV.   As a result, there is no analogue of the ``hadronic axion" window because we must always impose the cosmological constraint.


\section{Conclusions}
\label{sec:discussion}
In light of null results from WIMP direct detection experiments, there is considerable interest in exploring new regions of dark matter parameter space with a variety of different technologies. Several experimental topologies have recently been suggested for the detection of inelastically scattering dark matter in the mass range MeV - keV and the absorption of dark matter with mass greater than meV \cite{Knapen:2016cue, Hochberg:2016sqx, Schutz:2016tid, Hochberg:2016ajh, Essig:2016crl, Derenzo:2016fse, Essig:2015cda, Essig:2011nj, Graham:2012su} . This range of dark matter mass is heavily constrained from cosmology and astrophysics, particularly because CMB measurements constrain the entropy ratio between neutrinos and photons at temperatures at or below an MeV where these light dark matter particles could be thermally relevant. 

While there are stringent constraints on the interactions of electrons and dark matter, the nuclear couplings are relatively less constrained. In fact, the analog of the ``hadronic axion" window for dark matter-nucleon scattering would allow for enhanced direct detection cross-sections for mediators of mass around $\sim$ 100 keV. This window will soon be independently probed by CMB experiments like the Simons Observatory and CMB Stage IV~\cite{Abazajian:2016yjj}. If these experiments fail to see evidence for dark radiation, the dark matter will be outside the hadronic axion window and will have significantly lower scattering cross-sections, placing it near the solar neutrino floor. In addition to elastic scattering, light dark matter may also be absorbed. This scenario is not as constrained and there appears to be significant room for experimental progress. 

It should be noted that our cosmological constraints are not easily relaxed for light particles - they arise from the production of entropy in the universe around the time when neutrinos decouple and positrons annihilate. In light of CMB and BBN measurements, this entropy is difficult to hide. In the same vein, our results also rule out exotic electron-neutrino interactions that have been proposed to explain DAMA. Future CMB observations will place even more stringent constraints on light mediators and dark matter as measurements become sensitive to relics from thermalization prior to the QCD phase transition potentially back to the time of reheating.

\vskip23pt
\paragraph{Acknowledgements}
We thank Daniel Baumann, Vera Gluscevic, Peter Graham and Ben Wallisch for helpful conversations. We also thank Simon Knapen and Tongyan Lin for pointing out a numerical error in our cross-section limits.  SR was supported in part by the NSF under grants PHY-1417295 and PHY-1507160, the Simons Foundation Award 378243 and  the Heising-Simons Foundation grant 2015-038. 
\clearpage
\appendix
\section{Dark Radiation from Mediator Decay}
\label{app:Neff}
In this appendix, we will compute the predictions of $\Delta \Neff$ for a general dark sector in terms of $g_{\phi}$, $g_\varphi$ and $g_\star$: the mediator, decay products and visible degrees of freedom respectively.

We will be particularly interested in the situation where the mediator couples strongly to nuclei and is therefore in equilibrium above the QCD phase transition.  For completeness, we will only assume that the mediator was in equilibrium and freezes out at some temperature $T_F$. 

The simplest case arises when $m_\phi \ll 1$ eV and itself becomes dark radiation.  The conservation of comoving entropy from $T_F$ onward determines the relic abundance of $\phi$ in terms of $g_\star(T_F)$ and can be written as an additional contribution to $\Neff$, 
\beq\label{eq:decoupled}
\Delta \Neff^{\phi} = g_\phi \, \frac{4}{7} \,   \left( \frac{43}{4}\frac{1}{g_\star(T_F)} \right)^{4/3} \ .
\eeq
We know $g_\star(T)$ from the Standard Model and therefore we can predict $\Delta \Neff$, as shown in Figure~\ref{fig:DNeff} using $g_\phi = 1$.  However, due to the bounds from long range forces this situation does not arise in a regime where the dark matter could be observed directly.

When $m_\phi > 100$ eV a thermalized mediator must decay to another light field, $\varphi$, otherwise it would over-close the universe (decays to the Standard Model are excluded by the same arguments as in Section~\ref{sec:freezeout}).  This field must have $m_\varphi \ll 1$ eV so that we can hide its relic energy in dark radiation.  If the decay rate $\Gamma_{\varphi} \gg H(T=m_\phi)$, then the decay brings the $\varphi$ field into equilibrium and the relic abundance can be similarly predicted by the conservation of comoving entropy.  If $\Gamma_\varphi \gg H(T_F)$ then the new field was always in equilibrium and we can easily determine the additional contribution to $\Neff$, 
\beq\label{eqn:eq}
\Delta \Neff^{\varphi, {\rm early}} = \frac{g_{\varphi}}{g_\phi} \left(\frac{g_{\varphi} + g_\phi}{g_\varphi}\right)^{4/3} \times \Delta \Neff^{\phi}(T_F) \ .
\eeq
On the other hand, if $\varphi$ equilibrates at some temperature $m_\phi \ll T \ll T_F$, then the equilibration process conserves comoving energy (because both particles are relativistic), i.e. bringing $\varphi$ into equilibrium cools $\phi$ relative to the Standard Model.  Once in equilibrium, comoving entropy is conserved through the decay of $\phi$ to $\varphi$ at $T < m_\phi$.  The final abundance of $\varphi$ translates into a contribution to $\Neff$,
\beq\label{eqn:noeq}
\Delta \Neff^{\varphi} =   \left(\frac{g_{\varphi} + g_\phi}{g_\varphi}\right)^{1/3} \times \Delta \Neff^{\phi}(T_F) \ .
\eeq
There is still a relative enhancement to $\Delta \Neff$ but it is significantly reduced because of the initial cooling of the mediator temperature.  This is captured by the change in the power-law from $4/3 \to 1/3$.  If $\Gamma_\varphi  < H(T=m_\phi)$, the decay would happen out of equilibrium.  In such cases, the energy density in $\phi$ would grow relative to the Standard Model once it became non-relativistic and the contribution to $\Delta \Neff$ after its decay would be significantly enhanced.

The consequence for the mediator is that the best we can do to avoid observational constraints is to introduce a real scalar with $g_\varphi =1$  that comes into equilibrium with $\phi$ below $T_F$.  We will conservatively take the effective freeze-out temperature to be 300 MeV so that it lies above the QCD phase transition which produces
\beq
\Delta \Neff^\phi = 0.07 \qquad \qquad \Delta \Neff^\varphi = 0.09 \ .
\eeq
We could easily generate larger values by increasing $g_\varphi$ or $g_\phi$.  To avoid these predictions for large nucleon coupling would either require a very low reheat temperature $T_R < 300$ MeV or some out-of-equilbrium process that dilutes the mediator after the QCD phase transition but well before BBN.

\section{Dark Matter Abundance}
\label{app:abundance}

For the purposes of direct detection and cosmology, the simplified model discussed above provides a consistent and useful framework for future observations. However, from a model building point of view, one might wonder about the origin of the dark matter abundance. We have already established that the dark matter cannot get its abundance from thermal equilibrium with the Standard Model. While a non-thermal origin for the dark matter is certainly possible, it might be preferable to consider models where the dark matter abundance is set by thermal processes.  However, in our Equation~(\ref{eq:model1}) with $m_\phi > 10$ eV, there is simply no way to reduce the entropy carried by the dark matter without being in conflict with current data in one way or another (for example, from over-closure when $m_\phi , \md \gg 100$ eV).

A necessary condition is that we introduce a new light field, $\varphi$ with $m_\varphi \ll 1$ eV into which the dark matter can annihilate.  This can be accomplished by including the interaction ${\cal L} \supset\frac{1}{4} \lambda'  \chi^2 \varphi^2$.   Let us assume that $\varphi$ was thermalized at some temperature $T \gg \md$.  This is automatic if $\varphi$ is the same light field to which $\phi$ decays after $\phi$ freezes-out, which we will assume unless stated otherwise.  

At temperatures above the mass of the dark matter particle, the dark matter is produced by $\varphi +\varphi \to \chi +\chi$. When the temperatures drop below $\md$, the dark matter annihilates to $\varphi$ and leaves a relic abundance

\bea
\Omega_{\rm dm} &=& \sqrt{\frac{g_{\rm total}(m) \pi^2}{90}} \frac{x_f T_\varphi^3}{\langle \sigma v \rangle \rho_{\rm crit.}} \nonumber \\
&\approx& 0.2 h^{-2} \, \frac{x_f}{10} \left(\frac{\Delta\Neff^{\varphi}}{0.036} \right)^{3/4} \frac{10^{-27} {\rm cm}^3 \, {\rm s}^{-1}}{\langle \sigma v \rangle }
\eea
where $x_f = m / T_\varphi$ at freeze-out and $T_\varphi$ is the temperature of $\varphi$.  Since the final energy density in the mediator is determined by $T_\varphi$, we can write the results in terms of $\Delta \Neff$ given by 
\beq
\Delta \Neff^{\varphi} =   \left(\frac{g_{\varphi} + g_{\rm dm}}{g_\varphi}\right)^{\gamma} \times \Delta \Neff^{\varphi}(\lambda'=0) \ ,
\eeq
where $\gamma = \{ 4/3, 1/3 \}$ depending on the details of the thermal history, as explain in Appendix~\ref{app:Neff}, and $\Neff^{\varphi}(\lambda'=0)$ is  the contribution of $\varphi$ to $\Neff$ if it did not equilibrate with $\chi$ (i.e.~if $\varphi$ is thermalized by the decay of $\phi$, $\Delta \Neff^{\varphi}(\lambda'=0)$ are the values computed in Appendix~\ref{app:Neff}).  Although matching in the observed abundance in detail depends on the precise thermal history, qualitative agreement requires that
\beq\label{eq:abundancecoupling}
\lambda' \approx 10^{-7} \times \frac{\md}{1 \, {\rm MeV}} \ .
\eeq
We will not need a more precise estimate of the coupling as the purpose of this model is to provide order of magnitude estimates for current constraints and direct detection signatures.

The details of the thermal evolution are important when $m_\phi > 2 \md$ and the mediator can decay directly into the dark matter.  In Section~\ref{subsec:nucleon} we imposed the constraint that the dark matter could not be brought into equilibrium with $\phi$ through the decay which ultimately forces $g_N < 10^{-9}$ to avoid thermalization of $\phi$ altogether.  However, if dark matter has a thermal origin, thermalizing $\chi$ at 1 GeV would not over close the universe because it will dump its entropy into $\varphi$.  The coupling, $\lambda'$, is determined by the abundance of $\chi$ as in Equation~(\ref{eq:abundancecoupling}) and at $T \approx 1$ GeV, the rate of producing $\varphi$ from $\chi$ is 
\beq
\Gamma_\varphi \simeq \frac{1}{4\pi} \lambda'{}^2 T  \approx 10^{-15}  \text{ GeV} \times \frac{\md^2}{(1 \, {\rm MeV})^2} \ .
\eeq
For $\md > 5$ keV we have $\Gamma_\varphi > H(T\sim 1  \text{ GeV})$ and therefore $\phi$, $\chi$ and $\varphi$ would all be in equilibrium with the Standard Model at $T\sim 1$ GeV.  After $\phi$ freezes-out, $\phi$ decays to $\chi$ and $\chi$ annihilates to $\varphi$. The resulting contribution to $\Neff$ is 
\beq
\Delta \Neff^{\varphi} =  g_{\rm \varphi} \left(\frac{g_{\varphi} + g_{\rm dm}+g_\phi}{g_\varphi}\right)^{4/3} \times \Delta \Neff^{\phi} \geq  0.31 \ ,
\eeq
where we used $\Delta \Neff^{\phi}  \geq 0.071$.  We will not consider this possibility since it is nearly excluded at $2\sigma$ with current observations. 

The key difference between $m_\phi > 2 \md$ and $m_\phi < 2 \md$ is that in the latter case the dark matter ($\chi$) is not in thermal equilibrium with the Standard Model at 1 GeV. We can choose the coupling between $\chi$, $\phi$ and $\varphi$ so that $\varphi$ and $\chi$ are in equilibrium with each other only after $\phi$ has decoupled from the Standard Model - the entropy dump during the QCD phase transition allows $\phi$ to be colder than the Standard Model, diluting the contribution to dark radiation from the subsequent annihilation of $\chi$ into $\varphi$. 
 
\section{Dark Matter Sub-Component}
\label{app:nobullet}

The bullet cluster plays an important role in constraining $\lambda$ through the self-interaction of dark matter.  Unlike $g_f$, there are no lab-based bounds on $\lambda$ alone since it does not involve any Standard Model particles.  Cosmological or astrophysical bounds are the only way to place constraints.  However, these bounds do not apply to small sub-components of the dark matter that could be much more strongly interacting.  These sub-components could still be detected in the lab and are therefore still of interest experimentally.

\begin{figure}[h!]
\begin{center}
\includegraphics[width=0.49\textwidth]{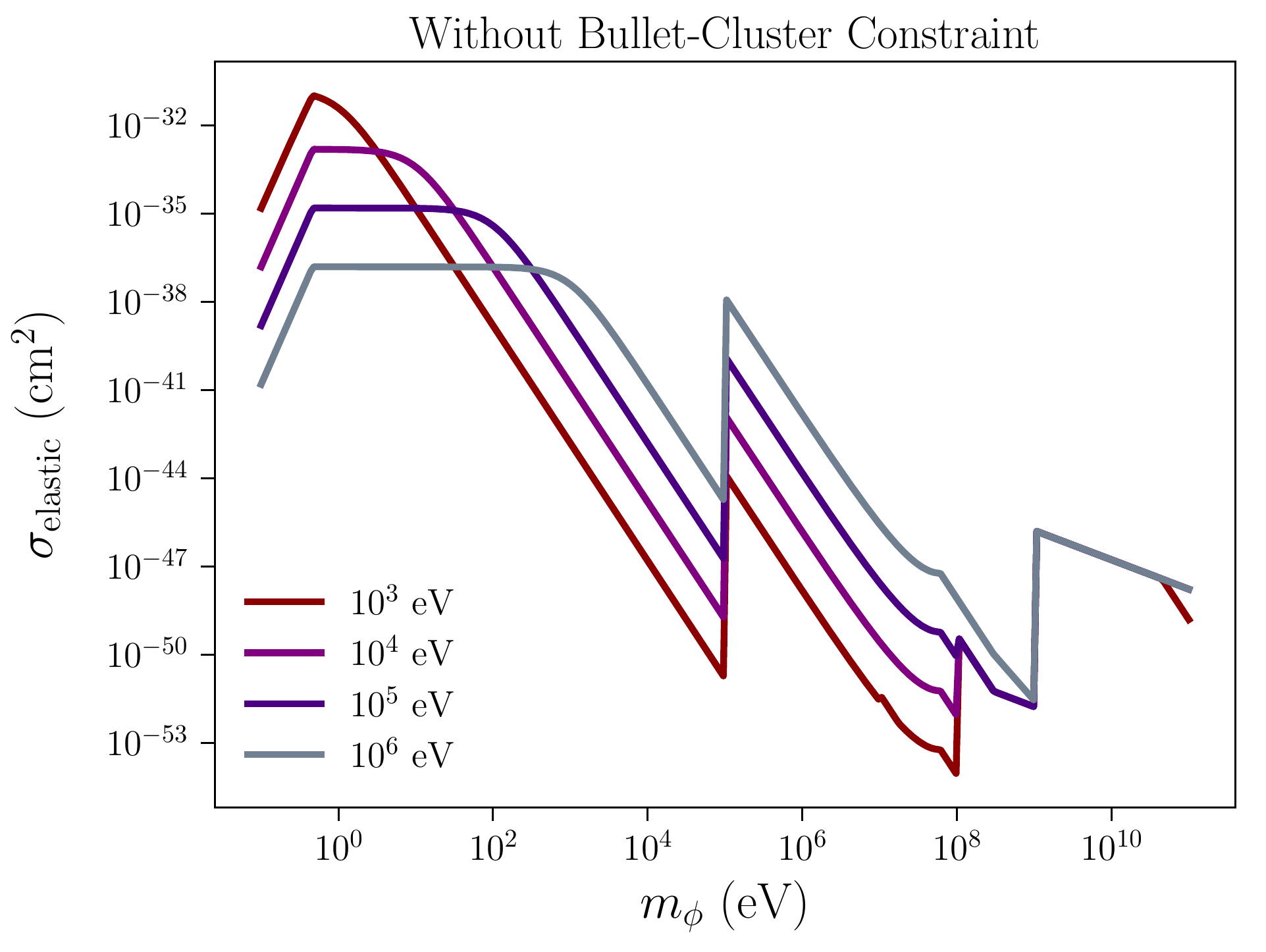}
\includegraphics[width=0.49\textwidth]{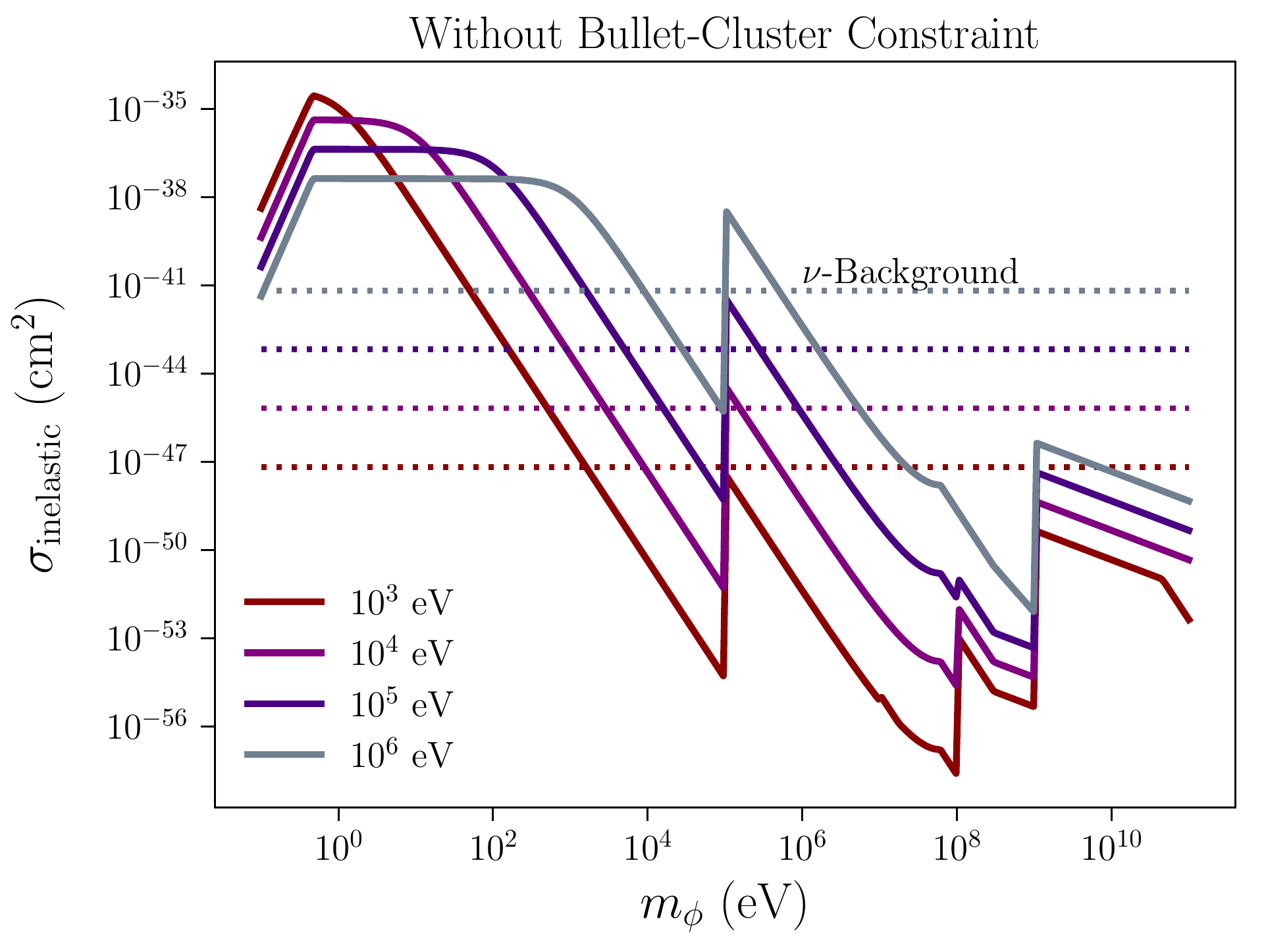}
\caption{{\it Left:}  Limits on dark matter-nucleon elastic cross-section assuming no bullet cluster limit, shown as a function of mediator mass for various dark matter masses assuming that we exclude $\Delta \Neff =0.09$ with future CMB data. {\it Right:}  Limits on the dark matter-nucleon inelastic cross-section assuming no bullet cluster limit, shown as a function of mediator mass for various dark matter masses, using $A=130$.  We also assume $\Delta \Neff =0.09$ is excluded with future CMB data.  The dotted lines show the neutrino background expected for the same momentum transfer as a dark matter part of the appropriate mass.  Note that without the bullet cluster, there are regions of $m_\phi$ and $\md$ that produce a conventional cosmology but produce cross-sections above the neutrino floor. }
\label{fig:nucleon_cross_nobullet}
\end{center}
\end{figure} 

When $m_\phi < v \, \md$, the bullet cluster constraint translates into roughly $\lambda < 10^{-7} (\md / {\rm MeV} )^{3/2}$.  Since the elastic and inelastic cross-section scale as $\lambda^2$, the cross-sections can be $10^{14-20}$ times larger than for the dominant component.  The cross-sections when imposing lab based constraints and imposing $\Delta \Neff <0.09$ are shown in Figure~\ref{fig:nucleon_cross_nobullet}.  Even with the cosmological constraints, we see that there is parameter space available significantly well above the neutrino floor at low mediator masses.  

\clearpage
\phantomsection
\addcontentsline{toc}{section}{References}
\bibliographystyle{utphys}
\bibliography{Refs}
\end{document}